\input harvmac




\def\+{^\dagger}
\def\d{{\delta}}
\def\e{{\rm e}}

\def \ep{\epsilon}

\def\H {{\cal H}}

\def \Q {{\hat Q}}
\def \P {{\hat  P}}

\def \k {\kappa}

\def \g {\gamma}
\def \G {\Gamma}
\def \del {\partial}

\def \a {\alpha}
\def \b {\beta}
\def \chi {\chi}
\def \s {\sigma}
\def \p {\phi}
\def \m {\mu}
\def \n {\nu}

\def \l {\lambda}

\def \td {\tilde }
\def \o {\omega}
\def \inv {^{-1}}
\def \ov {\over }

\def \fourth{{{1\over 4}}}

\def\d{{\delta}}
\def \hal{{{1\over 2}}}
\def \hs {{A^i_5}}
\def \bs {{B_{5i}}}
\def \Q {{\hat P}}
 \def \Q {{\hat Q}}
\def \R {{\cal R}}

\def\comment#1{}
\def \hal {{1\ov 2}}

\def\+{^\dagger}
\def\d{d}
\def\e{e}

\def \ep{\epsilon}

\def\H {{\cal H}}

\def \Q {{\hat Q}}
\def \P {{\hat P}}

\def \k {\kappa}

\def \g {\gamma}
\def \del {\partial}

\def \a {\alpha}
\def \b {\beta}
\def \chi {\chi}
\def \s {\sigma}
\def \p {\phi}
\def \m {\mu}
\def \n {\nu}

\def \l {\lambda}

\def \td {\tilde }

\def \o {\omega}
\def \inv {^{-1}}
\def \ov {\over }

\def \fourth{{{1\over 4}}}
\def \ha {{1\ov 2}}

\def\d{{\delta}}

\def\np {{  Nucl. Phys. }}
\def \pl {{  Phys. Lett. }}
\def \mpl {{ Mod. Phys. Lett. }}
\def \prl {{  Phys. Rev. Lett. }}
\def \pr  {{ Phys. Rev. }}

\def \Q {\hat Q}
\def \P {\hat P}


\def\comment#1{}
\def\fixit#1{}



\def\sqr#1#2{{\vcenter{\vbox{\hrule height.#2pt
         \hbox{\vrule width.#2pt height#1pt \kern#1pt
            \vrule width.#2pt}
         \hrule height.#2pt}}}}





\def\+{^\dagger}

\def\overleftrightarrow#1{\vbox{\ialign{##\crcr
     \leftrightarrow\crcr\noalign{\kern-0pt\nointerlineskip}
     $\hfil\displaystyle{#1}\hfil$\crcr}}}


\Title{
 \vbox{\baselineskip10pt
  \hbox{PUPT-1696} \hbox{SLAC-PUB-7452}\hbox{Imperial/TP/96-97/40}
  \hbox{hep-th/9704112}
 }
}
{
 \vbox{
  \centerline{ Intermediate Scalars and }
  \vskip 0.1 truein
  \centerline{the Effective String Model of Black Holes}
 }
}
\vskip -25 true pt
\baselineskip 16pt
\centerline{
 Igor R.~Klebanov,
 }
\centerline{\it Joseph Henry Laboratories,
Princeton University, Princeton, NJ 08544}

\centerline{ Arvind Rajaraman, }
\centerline{\it Stanford Linear Accelerator Center, Stanford, CA 94305 }

\centerline{ and}

\centerline{
 Arkady A.~Tseytlin\footnote
  {$^{\dagger}$}{Also at Lebedev Physics Institute, Moscow.} }
\centerline{\it Blackett
       Laboratory,  Imperial College,  London SW7 2BZ, U.K. }

\bigskip
\centerline {\bf Abstract}
\medskip
\baselineskip 12pt
We consider five-dimensional black holes modeled by D-strings bound to
D5-branes, with momentum along the D-strings.
We study the greybody factors for the non-minimally
coupled scalars which originate from
the gravitons and R-R antisymmetric tensor particles
polarized along the 5-brane, with one index along
the string and the other transverse to the string.
These scalars, which we call intermediate, couple to the
black holes differently from the minimal and the fixed scalars
which were studied  previously.
Analysis of their fluctuations around the black hole reveals
a surprising mixing between these NS-NS and R-R scalars.
We disentangle this mixing and obtain two decoupled scalar
equations. These equations have some new features, and we
are able to calculate the greybody factors only in certain limits.
The results agree with corresponding
calculations in the effective string model
provided one of the intermediate
scalars couples to an operator of dimension
(1,2), while the other to an operator of dimension (2,1).
Thus, the  intermediate scalars are
sensitive probes of the chiral operators in the effective string
action.

\Date{April 1997}

\noblackbox
\baselineskip 14pt plus 1pt minus 1pt


\def\jref{}  

\lref\ATT{A.A.~Tseytlin, {\it Mod.~Phys.~Lett.}~A11 (1996) 689,
hep-th/9601177.}

\lref\Ark{A. Tseytlin, {\it Nucl. ~Phys.}~B469 (1996) 51, hep-th/9602064.}

\lref\CGKT{
C.G.~Callan, Jr., S.S.~Gubser, I.R.~Klebanov and A.A.~Tseytlin,
\np B489 (1997) 65, hep-th/9610172;
I.R. Klebanov and M. Krasnitz, \pr D55 (1997) 3250, hep-th/9612051.}

\lref\CTT{M. Cveti\v c and  A.A.  Tseytlin,
\pl B366 (1996) 95, hep-th/9510097; \pr D53 (1996) 5619,
hep-th/9512031.}

\lref\CT{M.~Cveti\v c and A.A.~Tseytlin, {\it Nucl.~Phys.}~B478 (1996)
181, hep-th/9606033.}

\lref\CYY{M.~Cveti\v c and D.~Youm, {Nucl.~Phys.}~B476 (1996) 118,
hep-th/9603100\jref.}

\lref\CY{M.~Cveti\v c and D.~Youm, {\it Phys.~Rev.}~D53 (1996) 584,
hep-th/9507090; Contribution to `Strings 95', hep-th/9508058.}

\lref\Duff{M.J. Duff, H. Lu, C.N. Pope
Phys. Lett. B382 (1996) 73, hep-th/9604052.}

\lref\GKtwo{S.S.~Gubser and I.R.~Klebanov, Phys. Rev. Lett. 77
(1996) 4491, hep-th/9609076.}

\lref\GK{S.S.~Gubser and I.R.~Klebanov,
Nucl. Phys. B482 (1996) 173, hep-th/9608108.}

\lref\GR{I.S.~Gradshteyn and I.M.~Ryzhik, {\it Table of Integrals,
Series, and Products}, Fifth Edition, A.~Jeffrey, ed. (Academic Press:
San Diego, 1994).}

\lref\HM{G.~Horowitz and A.~Strominger, {\it Phys.~Rev.~Lett.}~77
(1996) 2368, hep-th/9602051.}

\lref\HP{G. Horowitz and J. Polchinski, hep-th/9612146.}

\lref\JP{J. Polchinski, \prl 75 (1995) 4724, hep-th/9510017.}

\lref\KT{I.R. Klebanov and A.A. Tseytlin, \np B475 (1996) 179,
hep-th/9604166.}

\lref\LS{L. Susskind, hep-th/9309145.}

\lref\maha{J.~Maharana and J.H.~Schwarz, {Nucl.~Phys.}~B390 (1993)
3, hep-th/9207016.}

\lref\GKT{ S.S.~Gubser, I.R.~Klebanov and A.A.~Tseytlin,
hep-th/9703040.}

\lref\at{A.A.~Tseytlin, {\it Nucl.~Phys.}~B475 (1996) 149,
hep-th/9604035.}

\lref\cm{C.G.~Callan and J.M.~Maldacena, {\it Nucl.~Phys.}~B472 (1996)
591, hep-th/9602043.}

\lref\dgm{S.~Das, G.~Gibbons and S.~Mathur, \prl 78 (1997) 417, hep-th/9609052}

\lref\dmII{S. Das and S.D. Mathur, {\it Nucl. ~Phys.}~B482 (1996) 153,
hep-th/9607149.}

\lref\dmI{S.R.~Das and S.D.~Mathur, {\it Phys.~Lett.}~B375 (1996) 103,
hep-th/9601152.}

\lref\dmw{A.~Dhar, G.~Mandal and S.~R.~Wadia, 
{\it Phys.~Lett.}~B388 (1996) 51, hep-th/9605234.}

\lref\dm{S.R.~Das and S.D.~Mathur, {\it Nucl.~Phys.}~B478 (1996) 561,
hep-th/9606185; hep-th/9607149\jref.}

\lref\dowk{F. Dowker, D. Kastor and J. Traschen, hep-th/9702109.}

\lref\fkk{S.~Ferrara and R.~Kallosh, {\it Phys.~Rev.}~D54 (1996) 1514,
hep-th/9602136; {\it Phys.~Rev.}~D54 (1996) 1525, hep-th/9603090;
S.~Ferrara, R.~Kallosh, A.~Strominger, {\it Phys.~Rev.}~D{52} (1995)
5412, hep-th/9508072.}

\lref\gibb{G.~Gibbons, {\it Nucl.~Phys.}~{B207} (1982) 337;
P.~Breitenlohner, D.~Maison and G.~Gibbons, {\it
Commun.~Math.~Phys.}~120 (1988) 295.}

\lref\gkk{G.~Gibbons, R.~Kallosh and B.~Kol, \prl77 (1996) 4992,
hep-th/9607108}

\lref\gkt{J.P.~Gauntlett, D.~Kastor and J.~Traschen, {\it
Nucl.~Phys.}~B478 (1996) 544, hep-th/9604179.}

\lref\gunp{S.S.~Gubser, November~1996, unpublished notes.}

\lref\hawk{S. Hawking and M. Taylor-Robinson, hep-th/9702045 .}

\lref\hmf{{\it Handbook of Mathematical Functions}, M.~Abramowitz and
I.A.~Stegun, eds. (US Government Printing Office, Washington, DC,
1964) 538ff.}

\lref\hms{G.~Horowitz, J.~Maldacena and A.~Strominger, {\it
Phys.~Lett.}~B383 (1996) 151, hep-th/9603109.}

\lref\hrs{E.~Halyo, B.~Kol, A.~Rajaraman and L.~Susskind,
hep-th/9609075\jref; E.~Halyo, hep-th/9610068\jref.}

\lref\HS{G.~Horowitz  and A.~Strominger, Nucl. Phys. B360 (1991) 197.}

\lref\jpTASI{J.~Polchinski, hep-th/9611050\jref.}

\lref\juanI{J.~Maldacena, Rutgers preprint RU-96-102,
hep-th/9611125\jref.}

\lref\juan{J.~Maldacena, {Nucl.~Phys.}~{ B477} (1996) 168,
hep-th/9605016.}

\lref\kaaa{R.~Kallosh, A.~Linde, T.~Ort\'in, A.~Peet and A.~Van
Proeyen, {\it Phys.~Rev.}~D{46} (1992) 5278.}

\lref\km{I.R.~Klebanov and S.D.~Mathur, Princeton preprint PUPT-1679,
MIT preprint MIT-CTP-2610, hep-th/9701187.\jref}

\lref\kr{B.~Kol and A.~Rajaraman, Stanford preprint SU-ITP-96-38,
SLAC-PUB-7262, hep-th/9608126\jref.}

\lref\ktI{I.R.~Klebanov and A.A.~Tseytlin, Princeton preprint
PUPT-1639, {\it Nucl.~Phys.}~B479 (1996) 319, hep-th/9607107.}

\lref\kt{I.R. Klebanov and A.A. Tseytlin, \np B475 (1996) 165,
hep-th/9604089.}

\lref\lu{J.X.~Lu, {\it Phys.~Lett.}~B313 (1993) 29, hep-th/9304159.}

\lref\MSII{J.M.~Maldacena and A.~Strominger, Santa Barbara preprint
UCSBTH-97-02, hep-th/9702015.}


\lref\mst{J.M.~Maldacena and A.~Strominger, {\it Phys.~Rev.~Lett.}~77
(1996) 428, hep-th/9603060.}

\lref\ms{J.M.~Maldacena and L.~Susskind,
{\it Nucl.~Phys.}~B475 (1996) 679, hep-th/9604042.}

\lref\myers{C. Johnson, R. Khuri and R. Myers,
Phys. Lett. B378 (1996) 78, hep-th/9603061.}

\lref\schw{J.H.~Schwarz, {\it Nucl.~Phys.}~B226 (1983) 269.}

\lref\sv{A.~Strominger and C.~Vafa, {\it Phys.~Lett.}~B379 (1996) 99,
hep-th/9601029.}

\lref\StVa{ A. Strominger and C. Vafa,  Phys. Lett. { B379} (1996) 99,
hep-th/9601029.}

\lref\ClMl{ C.G. Callan and J. Maldacena, Nucl. Phys. { B472} (1996) 591,
hep-th/9602043. }

\lref\HS{ G. Horowitz and A. Strominger,  Phys. Rev. Lett. { 77} (1996) 2368,
hep-th/9602051.}

\lref\HMS{ G. Horowitz, J. Maldacena and A. Strominger,
 Phys. Lett. { B383} (1996) 151,
hep-th/9603109.}

\lref\MSIII{J. Maldacena and L. Susskind,  Nucl. Phys. { B475} (1996) 670,
hep-th/9604042.}

\lref \TTS{A.A. Tseytlin, \mpl { A11} (1996) 689, hep-th/9601177.}

\lref\DMW{ A. Dhar, G. Mandal and S.R. Wadia,  Phys. Lett. { B388} (1996)
51, hep-th/9605234. }

\lref\DM{ S.R. Das and S.D. Mathur,  Nucl. Phys. { B478} (1996) 561,
hep-th/9606185.}

\lref\DMIII{ S.R. Das and S.D. Mathur, Nucl. Phys. { B482} (1996) 153,
hep-th/9607149.}

\lref\MSI{ J. Maldacena and A. Strominger, Phys. Rev. { D55} (1997) 861,
hep-th/9609026.}

\lref\K{ I.R. Klebanov, hep-th/9702076.}

\lref\KKI{ I.R. Klebanov and M. Krasnitz,  hep-th/9703216.}

\lref\KVK{ E. Keski-Vakkuri and P. Krauss,  hep-th/9610045. }

\lref\KM{ I.R. Klebanov and S.D. Mathur, hep-th/9701187.}

\lref\MSII{ J. Maldacena and A. Strominger, hep-th/9702015.}

\lref\SWH{ S.W. Hawking and M.M. Taylor-Robinson,
 hep-th/9702045.}

\lref\Kol{ B. Kol and A. Rajaraman, hep-th/9608126.}

\lref\NRT{ J. Maldacena,  hep-th/9611125.}

\lref\Page{ D.N. Page,  Phys. Rev. { D13} (1976) 198.}

\lref\WGU{ W.G. Unruh, Phys. Rev. { D14}
(1976) 325.}

\lref\AbSt{ M. Abramowitz and I.A. Stegun, {\it Handbook of Mathematical
Functions}, National Bureau of Standards New York: Wiley (1964).}



\newsec{Introduction}
There has been much progress recently in describing the microstates
 of black holes through D-brane physics. The Bekenstein-Hawking entropy
of certain extremal and near-extremal black holes can be understood
through the counting of D-brane microstates \refs{\StVa,\ClMl,\HS,\HMS,\MSIII}.
Furthermore, the Hawking radiation and the semi-classical
absorption were shown in many
cases to agree with the calculation of the corresponding process
in the D-brane picture. This was demonstrated for the charged
black holes in four and five dimensions that are described by effective
string models
\refs{\DM,\DMW,\GK,\DMIII,\MSI,\GKtwo,\KVK,\KM,\hawk,\dowk,\MSII},
as well as for the extremal threebranes that admit a direct D-brane
description \refs{\K,\GKT}.

The results mentioned so far refer to minimally coupled scalar fields.
Not all scalars, however, are minimally coupled. There are other scalars which
couple to the non-trivial vector backgrounds. Examples of these are the
`fixed' scalars considered in \refs{\Kol,\CGKT}, which have
different cross-sections from the minimally coupled scalars.
In the $D=5$ black hole background there are two specific  fixed scalars,
which mix with each other and with the
gravitational perturbations \CGKT.
Recently,  the complexities of this mixing were disentangled
in \KKI. The greybody factors calculated from the diagonalized
equations of motion were found
to be of the form obtained  earlier in \CGKT: in the effective string
model such greybody factors are reproduced by operators of dimension
$(2,2)$. This poses a puzzle, since the effective string action
derived in \CGKT\
also contains couplings to dimension $(1,3)$ and $(3, 1)$ operators
which produce greybody factors of a different form.
Thus, it is of special interest to study other situations in which
chiral operators appear in the effective string couplings.
This  will be  the subject of the present paper.

We will be concerned with
yet a third type of scalars, which we call intermediate,
first considered in \CGKT. This type is
different from both the minimally coupled and the fixed scalars.
The intermediate scalars originate from the fields $\hs$ (denoted by
$h_{5i}$ in \CGKT)  and
$B_{5i}$, i.e. the gravitons and the R-R 2-form particles
polarized along the 5-brane, with one index pointing along
the string and the other transversely to the string.
In this paper, we will calculate
the semi-classical absorption cross-sections
of the intermediate scalars and compare them with the effective string model
predictions.

In Section 2,
after presenting the setup, we derive the classical
equations of motion for the intermediate
scalars in the $D=5$ black hole background (an alternative derivation
based on the 6-dimensional theory will be presented in the Appendix).
This turns out to be quite nontrivial due to a mixing between $\hs$
and $B_{5i}$.
In Section 3 we propose a coupling for these scalars
in the effective string model of the black hole. Part of this
coupling term is not present in the  standard  Nambu-type  D-string
 action.
 It turns out that requiring the scalars to couple to operators of
a given dimension on the world sheet is a very restrictive guiding
principle. We find that the necessary operators are of dimensions
(1,2) and (2,1) and then calculate
the resulting cross-sections as predicted by the effective string model.

Finally,  in Section 4  we compare the
absorption cross-sections derived by semi-classical
considerations to the cross-sections
predicted by the string. The classical
equations of motion are complicated and we are only able to solve for
the cross-sections in various limits. In
every case that we can treat analytically,
there is exact agreement between the
semi-classical gravity and the effective string.
This is  evidence that the effective string model
reproduces the dynamics of the intermediate scalars. However, our inability
to solve for the general semi-classical greybody factor 
leaves the question
of the complete agreement open.

\newsec{ Derivation of the Equations of Motion}
\bigskip

 As in \CGKT\  we start with the action of the  $D=10$
 type IIB  supergravity
 reduced  to 5 dimensions. The relevant part of it is
 \eqn\actit{
S_5 =
 {1 \ov 2 \k_5^2} \int d^5 x \sqrt g \bigg[ R  - {4\ov 3}(\del_\m \p_5)^2
 -  { 1 \ov 4} G^{pl} G^{qn}(\del_\m G_{pq} \del^\m G_{ln}
 + \sqrt {G}  \del_\m B_{pq} \del^\m B_{ln} )
}  $$ -\
 \fourth e^{-{4\ov 3} \p_5   }G_{pq} F^{p}_{\m\n}  F^{q}_{\m\n}
- \fourth e^{{2\ov 3}\p_5 } \sqrt {G} G^{pq}  H_{\m\n p } H_{\m\n q }
-  {1\ov 12} e^{{4\ov 3} \p_5 } \sqrt {G}  H^2_{\m\n\l} + ...
 \bigg]   \ ,  $$
where $\m,\n,...= 0,1,2,3,4;\  p,q,...=5,6,7,8,9$. \
 $\p_5$ is the  5-d dilaton and $G_{pq}$ is the metric of
internal 5-torus,
$$\p_5 \equiv \p_{10} - \fourth G = \p_6 - \ha \l
\ , \qquad G= \det G_{pq}\ ,
$$
and $B_{pq}$ are the internal components of the R-R 2-tensor.
$F^{p}_{\m\n}$ is the field strength of the Kaluza-Klein vectors
$A^{p}_\m$.
It will be crucial in what follows that
$H_{\m\n p}$ and $H_{\m\n\l}$ are
  given explicitly by (see, e.g., \maha)
  \eqn\dewf{
  H_{\m\n p} =   F_{\m\n p}  - B_{pq} F^{q}_{\m\n} \ ,
  \ \ \ \  F_p = dB_p \ , \ \ \ F^{p} = d A^{p} \ ,
    }
  $$
  H_{\m\n\l} = \del_\m B_{\n\l} - \hal  A^{p}_\m F_{\n\l p}
  - \hal  B_{\m p} F^{p}_{\n\l} + {\rm cyclic\  permutations} \ ,  $$
  where $B_{\m p }$ and $B_{\m\n}$  differ from
   the  $D=10$  components  of the R-R
2-form field by terms proportional to
$A^{p}_\m$.
 The  `shifts' in these field strengths vanish for the $D=5$ black hole
 backgrounds which correspond to bound states of RR-charged
5-branes and strings with momentum flow.
 For such black holes, $B_{pq}=0$,
 the vector fields $A^{p}$ and $B_p$ have electric backgrounds,
while $H_{\m\n\l}$ has a magnetic one
 (we shall  assume that
the electric  charges $Q_{Kp}$ and $Q^p$
 corresponding to the vectors $A^{p}$ and $B_p$
have only the $p=5$ component).
However, in general the field strength shifts in
\dewf\ are  important for the discussion of perturbations.
We will argue, in fact, that while the shift in $H_{\m\n\l}$
does not contribute in the present case, the shift in $H_{\m\n  p }$
will lead to a mixing between perturbations of $G_{pq}$ and $B_{pq}$
for $p=5$ and $q=i$ \
($5$ is the direction of the string and $i=6,7,8,9$
label the directions of $T^5$ orthogonal to the string).

The 5-dimensional charged black hole metric is  \refs{\TTS,\CYY,\HMS}
\eqn\met{
ds^2_5 = g_{mn} dx^m dx^n + ds^2_3=
 - h \H^{-2} dt^2 + h\inv \H  dr^2   +  r^2  \H   d\Omega_3^2  \ , }
  $$ h=1 - { r^2_0 \ov r^2}\  , \ \  \ \ \H\equiv (H_n H_1 H_5)^{1/3}\ , \ \ \
\ \
  \sqrt g = r^3 (H_n H_1 H_5)^{1/3} \ ,  $$
 $$
H_1 = 1 + {\Q\ov r^2} \ , \ \ \ H_5 = 1 + {\P\ov r^2} \
 , \ \ \ H_n = 1 + {\Q_K \ov r^2}  \ ,  $$
where $ \Q = \sqrt { Q^2 + \fourth r^4_0}  - \ha r_0^2$, etc.
The background values  of the internal
metric and the  dilaton  are (see \CGKT\ for more details)
\eqn\torr{
(ds^2_{10})_{T^5} =  G_{pq} dx^p dx^q =
e^{2\n_5} dx_5^2 +  e^{2\n }
(dx^2_6 + dx^2_7 + dx^2_8 + dx^2_9) \  ,   }
$$
\n_5 = -2 \p_5 \equiv  \l  \ , \ \ \ \ \ \ \    e^{2\l} =
  {H_n \ov (H_1H_5)^{1/2} } \ . $$
It is useful to choose the
 following parametrization for the full (background plus perturbation)
 internal metric
\eqn\maay{
 G_{pq}
 =
e^{2\n}  \pmatrix{    e^{2\l - 2\n }   + e^{2\n}   A^i_5 A^i_5   &     A^i_5
  \cr      A^j_5
 &     \delta_{ij}  \cr} \ ,  \ \ \    \ \ \     \sqrt G = e^{\l + 4\n} \ ,
 }
$$
 { G}^{pq}  = e^{-2\l}  \pmatrix{ 1  &  -  A^i_5
 \cr
   -     A^j_5
&  e^{2\l- 2\n} \delta^{ij} +  A^i_5 A^j_5 \cr}\  ,
 $$
 For the present  discussion of the `off-diagonal' perturbations
 the fluctuations of $\p_5$, as well as those of $\sqrt G$,
 can be ignored.
Therefore,  we concentrate on the dependence on $\hs$ and $\bs$
 and do not keep track of other scalar
perturbations which  were  already  discussed in \CGKT.

The $D=5$ scalars $\hs$ and $\bs$ originate from the
$M=5$ components of the KK vector $A^i_M$  and the vector component
$B_{Mi}$ of the R-R 2-tensor in  type IIB supergravity reduced to  6
dimensions. An alternative derivation of the equations for the
$\hs$ and $\bs$ perturbations,
which directly uses the  $D=6$ theory, will be presented in the Appendix.

The relevant  terms  in the $D=5$ action
are\foot{The $\m,\n$ indices are always contracted using
the curved  5-dimensional metric and assuming that
$F_{\m\n}F_{\m\n}\equiv F_{\m\n}F^{\m\n}$, etc.  The repeated $i,j$-indices are
summed  with $\d_{ij}$ with no extra factors (all factors in $5,i$ directions
are given explicitly).}
\eqn\acit{
S_5 =
 {1 \ov 2 \k_5^2} \int d^5 x \sqrt g \bigg[
 -  { 1 \ov 2}  e^{-2\l + 2\n } \del_\m \hs  \del^\m  \hs
 -  { 1 \ov 2}  e^{-2\l+ 2\n } \del_\m \bs  \del^\m  \bs
} $$ -\
 \fourth e^{{2\ov 3} \l + 2\n  }
 [ F^{i }_{\m\n}  F^{i}_{\m\n}
  + 2 F^{5 }_{\m\n}  F^{i}_{\m\n}    \hs +
 F^{5 }_{\m\n}  F^{5}_{\m\n} ( e^{2\l - 2\n } + \hs \hs ) ] $$
 $$
 - \ { 1\ov 4}  e^{-{4\ov 3}\l + 4\n}
\big [   H_{\m\n 5 } H_{\m\n 5 } -
  2    H_{\m\n i } H_{\m\n 5 } \hs
   +  H_{\m\n i } H_{\m\n j} ( e^{2\l - 2\n}
\d^{ij} +  A^i_5 A^j_5 ) \big ] \bigg]\ ,
   $$
where
$$
 H_{\m\n 5 } =  F_{\m\n 5}  - B_{5i} F^{i}_{\m\n} \ , \ \ \ \
  H_{\m\n i } =  F_{\m\n i}  +  B_{5i} F^{5}_{\m\n} \ .
$$
Here only $F_{\m\n 5}$ and $F^{5}_{\m\n}$ have background values,
which we denote by $\tilde F$,
$$
e^{{8\ov 3} \l  }   \sqrt g   (\tilde F^{5})^{0r}   =2 Q_K  \ , \
\ \ \ \
e^{-{4\ov 3}\l + 4\n}
\sqrt g   (\tilde F_{ 5})^{0r}    =2  Q  \ ,
$$
 so that
\eqn\qeq{\tilde F^{5}_{\m\n}  =   { Q_K \ov Q} e^{-4\l + 4\n } \tilde
F_{\m\n 5} \ . }
As a result, we may integrate out $F_{\m\n i}$ or all of $H_{\m\n i }$
 easily.
 This gives\foot{The
$H_i H_i hh$ term is of subleading order being quartic in the
fluctuations.}
\eqn\abo{
 - \fourth e^{-{4\ov 3}\l + 4\n}
 \big( -   e^{-2\l + 2\n } \td  F_{\m\n 5 } \td F_{\m\n 5 } \hs \hs  \big)
 \ . }
To show this  it is crucial that
 $\td F^{5}_{\m\n}$  has only the electric component and depends
only on $r$,
 and  that the scalar perturbations depend only on $r$ and $t$.
 This is  similar to what happens  in the  fixed scalar
case \refs{\Kol,\CGKT}.

 The mixing
  that  contributes  a new term
is $  \td F_{\m\n 5} B_{5i} F^{i}_{\m\n}  $
   which   comes from the  $H_{\m\n 5 }^2$
   term.\foot{One way to  see why the
 mixing terms inside $H_{\m\n\l}$ in  \actit\ and \dewf\
do not contribute is to
 dualize  $B_{\m\n}$ into a vector, $V_\m$.
The resulting terms in the action will  have the following structure:
$\int d^5x [  - \fourth \sqrt g e^{-{4\ov 3} \p_5 }
G^{-1/2}  F^2_{\m\n} (V) +
 \ep^{\m\n\l \s\k}  V_\m  F_{\n\l p} F^{p}_{\s\k}] . $
 The three
 vectors, $V_\m$, $A_{\m5}$, $A^{5}_\m$, have electric backgrounds
with charges $P, Q, Q_K$ respectively.
 The  trilinear Chern-Simons-type  term  produces a  non-zero
 contribution in the gaussian approximation
 only if the two  fluctuation   fields
 have indices different from $0$ and $r$,
 which are the directions of the electric background of the third field
 in the product.
 This means that the Chern-Simons-type term
 does not mix  the  `electric' perturbations of the fields,
 but it is the `electric' perturbations of the vectors that
  couple to the  off-diagonal scalars  we discuss.}
  The relevant vector-scalar terms are
 $$ -\
 \fourth e^{{2\ov 3} \l   } e^{2\n}
 \big[ F^{i }_{\m\n}  F^{i}_{\m\n}
  + 2 \td F^{5 }_{\m\n}  F^{i}_{\m\n}  \hs +
 \td F^{5 }_{\m\n}  \td F^{5}_{\m\n}  \hs \hs
  - 2 e^{-2\l + 2\n }   \td F_{\m\n 5} B_{5i} F^{i}_{\m\n} \big]\ .  $$
 It remains to integrate out $ F^{i}_{\m\n} $.
One should actually integrate over the corresponding gauge potential,
 but since the background is electric and static,
 and the scalars depend only on
 $r$ and $t$,  this is equivalent to just solving for the  field strength.

Adding  the  $\hs\hs$ term already obtained in \abo, we get
  $$  -
 \fourth e^{-{2\ov 3} \l  + 2\n}
 \bigg( - \big[\td F^{5}_{\m\n}  \hs  - e^{-2\l + 2\n }   \td F_{\m\n 5}
B_{5i}\big]^2 $$  $$
 +\
 \td F^{5 }_{\m\n}  \td F^{5}_{\m\n}  \hs \hs
  -   e^{-4\l + 4\n } \td  F_{\m\n 5 } \td F_{\m\n 5 } \hs \hs  \bigg) .  $$
 We can simplify this  expression using \qeq:
 $$
 \fourth e^{-{2\ov 3} \l  + 2\n   }
 \bigg(
    e^{-4\l + 4\n} \td  F_{\m\n 5 } \td F_{\m\n 5 }    ( \hs \hs  + \bs \bs )
-     2 e^{-2\l + 2\n }  \td F^{i}_{\m\n} \td F_{\m\n 5} \hs B_{5i}
 \bigg) $$
 \eqn\gins{
 = \fourth e^{-{14\ov 3} \l  + 6\n   }  \td  F_{\m\n 5 } \td F_{\m\n 5 }
 \bigg(
       \hs \hs  + \bs \bs \
-     2 { Q_K \ov Q} e^{-2\l + 2\n }    \hs B_{5i}
 \bigg) \ . }
The novelty is the mixing term in the brackets
 $$ -     2  {  Q_K \ov Q} e^{-2\l + 2\n }    \hs B_{5i}
 = - 2 { Q_K  H_1  \ov Q H_n }\hs B_{5i} \ ,   $$
 which  is thus  present  for arbitrary non-vanishing
 values  of $P$, $Q$ and $Q_K$.

Remarkably, the  full $\hs,\bs$ scalar  action
with the kinetic terms included
 can be diagonalized in terms of the fields  $\xi_i$ and  $\eta_i$ defined by
\eqn\qqq{  \eta_i = {1\ov \sqrt 2 } ( \hs + \bs ) \ , \ \ \ \  \ \
\xi_i = {1\ov \sqrt 2 } ( \hs - \bs ) \ .
}
With these definitions,
 \eqn\ait{
S_5 =
 {1 \ov 2 \k_5^2} \int d^5 x \sqrt g \bigg(
 -  { 1 \ov 2}  e^{-2\l + 2\n } [(\del_\m \xi_i)^2  +(\del_\m \eta_i)^2]
 } $$ + \ \fourth e^{-{14\ov 3} \l    + 6\n} \td  F_{\m\n 5 } \td F_{\m\n
5 }
 \big[ (1 +   {  Q_K \ov Q} e^{-2\l+2\n } )
        \xi_i ^2
 +   (1  -   {  Q_K \ov Q} e^{-2\l+2\n} )  \eta_i^2 \big] \bigg)  \ . $$
Rescaling  the fields to eliminate the  background-dependent
factors $e^{-2 \l+2\n}$
in the kinetic parts, we arrive at the following decoupled equations
(we shall use the same notation, $\xi_i$ and $\eta_i$, for the redefined
fields, $e^{-\l + \n } \xi_i$ and $ e^{-\l + \n } \eta_i$)
 \eqn\kgoo{ \eqalign{&
 \left[h r^{-3}  {d\over dr} (  h r^3 {d\over dr} )
 + \omega^2      H_1 H_5  H_{n }  - M_\xi     \right] \xi_i
\ = 0 \ , \cr
& \left[h r^{-3}  {d\over dr} (  h r^3 {d\over dr} )
 + \omega^2      H_1 H_5  H_{n }  - M_\eta    \right] \eta_i
\ = 0 \ , \cr
}}
 where
 \eqn\masses{
 M_\xi = M_{\l-\n} +   M_+ \ , \ \ \ \ \
 M_\eta = M_{\l-\n} +   M_- \ , }
 $$ M_{\l-\n}  =
h e^{\l-\n} r^{-3}  {d\over dr} ( r^3 h {d\over dr} e^{-\l+\n}) \ ,  $$
$$ M_\pm
=  { 4Q^2  \ov r^6   H_1^2 } ( 1 \pm  {Q_K  H_1  \ov Q H_n  } ) h
\ . $$
Somewhat surprisingly,  all the
  dependence on $P$ disappears  from
the  ``mass"  terms  since
$$  e^{-\l + \n}   = \big(H_1/ H_n\big)^{1/2}  \ ,  $$
so that
$$
M_{\l-\n}  ={ (\Q_K-\Q)(r^2-r^2_0)\over
 r^4 (r^2 +\Q_K)^2 (r^2 + \Q)^2 }
 \bigg[ (\Q + 3\Q_K + 2 r^2_0)r^4 $$
$$  + \  (4\Q\Q_K + \Q r^2_0  - \Q_K r^2_0) r^2  - 2\Q \Q_K r^2_0
\bigg]\
  . $$
In  the extremal limit, $r_0=0$,\ $\Q=Q, \ \P=P,\ \Q_K=Q_K$,
the resulting   ``mass" terms  are found to be
$$
M_\xi =   { 8 Q^2 Q^2_K  + 8 QQ_K (Q+Q_K) r^2 + (3Q^2 + 2 QQ_K
+ 3 Q^2_K) r^4
\over
 r^2 (r^2 +Q_K)^2 (r^2 + Q)^2 }
 \ , $$
 \eqn\rew{
 M_\eta  = {3 (Q-Q_K)^2  r^2 \ov  (r^2 + Q_K)^2 (r^2 + Q)^2  }
\ . }
 They have the following asymptotics
 $$
 r\to 0: \ \ \ \ \   M_\xi = { 8 \ov r^2} \ , \  \ \ \   \ M_\eta = 0 \ , $$
 $$
 r\to \infty : \ \ \ \ \  M_\xi =
{ 3Q^2 + 2 QQ_K  + 3 Q^2_K  \ov r^6} \ , \ \  \ \
 M_\eta = { 3 (Q-Q_K)^2 \ov r^6 } \ .  $$
At the horizon $\eta_i$ behaves as the $l=0$ partial wave of
a minimally coupled scalar. $\xi_i$, on the other hand,
behaves as the $l=2$ partial wave,
which is the behavior previously encountered for the fixed scalars
\refs{\CGKT,\KKI}.
The expressions \rew\  can  be simplified if $ Q \gg Q_K$,
\eqn\yty{
M_\xi =   { Q^2(8  Q^2_K  + 8  Q_K r^2 + 3 r^4)
\over r^2 (r^2 +Q_K)^2 (r^2 + Q)^2 } \ , }
 $$
 M_\eta  = {3 Q^2  r^2 \ov  (r^2 + Q_K)^2 (r^2 + Q)^2  } \ .  $$
Note that, for  $Q_K=0$,
\eqn\kgooo{
M_\eta = M_\xi=   { 3 Q^2
\over
 r^2  (r^2 + Q)^2 } \ . }
 Thus, as one switches on $Q_K$,
there is a remarkable jump from the $l=1 $ to the $l=0$
or $l=2$ behaviors near the horizon.

\newsec{
Absorption in the Effective String Model}

In the previous section we found a surprising mixing between
the off-diagonal components of the  Kaluza-Klein  scalars
 $\hs$ and
the internal components $B_{5i}$ of the  R-R 2-tensor.
In this section we
discuss this mixing from the effective string point of view, and
show what it implies about the greybody factors.

First, we have to write down the
lowest-dimension couplings to the effective string for
the fields in question.
In \CGKT\  the scalar fields $\hs$ were  included, but the components
$B_{5i}$ of the R-R field were  omitted. In fact,
as the discussion of the gravitational perturbations shows,
these two field mix and one should keep both of them.
The simplest assumption that one usually makes is that the effective
string action is the same as the D-string action with a rescaled tension.
The necessary terms in the action are then
$$ S= -T_{\rm eff} \int d^2 \sigma \ \big( \sqrt {- \g} -  \hat B_{05} \big)
\ ,
$$
where
$$ \g_{ab} = G_{\m\n} (X) \del_a X^\m \del_b X^\n\ ,\qquad
\hat B_{ab} = B_{\m\n} (X) \del_a X^\m \del_b X^\n \ .
$$
The leading order couplings are found to be
$$ -{T_{\rm eff}\over 2} [ \hs (\partial_+ + \partial_-) X^i
+ B_{5i} (\partial_+ - \partial_-) X^i] =
-{T_{\rm eff}\over \sqrt 2} [ \xi_i \partial_- X^i
+ \eta_i \partial_+ X^i ]\ ,
$$
where the same mixtures of the fields naturally emerge
as the ones needed in the
effective field theory (GR)  calculation, \qqq.
We see that these mixtures couple to operators of dimension $(0, 1)$
and $(1, 0)$ respectively.
Clearly, these couplings do not contribute to absorption.
Expanding further we find the term
$$ -{T_{\rm eff}\over 4} \hs\ [
\partial_- X^i (\partial_+ X)^2 +
 \partial_+ X^i (\partial_- X)^2 ]
\ ,$$
whose natural supersymmetric completion is
$$ -{T_{\rm eff}\over 4}\  \hs\  [\
\partial_- X^i\ T^{tot}_{++}\ +\
 \partial_+ X^i\ T^{tot}_{--}\ ]
\ ,$$
with $T^{tot}$ including the fermionic contribution as well.
It is interesting that,
using this coupling in the case $Q_K=0$, we find the greybody
factor which {\it exactly} agrees with the GR result.
So, for $Q_K=0$ (the non-chiral case) we may just use the coupling
stated in \CGKT\ and arrive at complete agreement with the
semi-classical calculation.

The structure of the action is less clear for $Q_K>0$.
While we do not readily see a cubic coupling for $B_{5i}$, we will
add it by hand to enforce the principle that $\xi_i$
and $\eta_i$ couple to operators of a given dimension.
With this assumption,
the terms that arise in the effective string action are
\eqn\actionterm{
\delta S=-{T_{\rm eff}\sqrt 2\over 4}
\int d^2 \sigma \ \big[\ \eta_i\  \partial_- X^i\  T^{tot}_{++}\  +\
 \xi_i\  \partial_+ X^i\  T^{tot}_{--}\  \big]
\ .
}

Using the action \actionterm,
let us now derive the effective string absorption cross-section for
$\eta_i$.
The absorption cross-section is due to processes
$\eta_i\rightarrow L+L+R$ and $\eta_i+L \rightarrow L+R$\
($L$ and $R$ stand for the
left-moving  and right-moving modes on the string).
The matrix element between properly normalized states,
including the kinetic term  normalization factor
$\kappa_5 \sqrt 2$ for $\eta_i$ (see \ait), is
found to be
\eqn\MatrixEl{
   {2\kappa_5\over \sqrt {T_{\rm eff}}}
\sqrt{ q_1 p_1 p_2\over \omega}\  \ .
  }
Adding up the absorption rates for the two
processes gives (see  \CGKT\ for
details of analogous cross-section computations)
\eqn\llr{ \eqalign{&
{3\kappa_5^2 L_{\rm eff} \over 2\pi T_{\rm eff} }
{1\over 1- \e^{-{\omega\over 2 T_R}} }
       \int_{-\infty}^\infty d p_1 d p_2 \,
        \delta\left( p_1 + p_2 -{\omega\over 2} \right)
        {p_1 \over 1 - \e^{-{p_1\over T_L}}}
        {p_2 \over 1 - \e^{-{p_2\over T_L}}}\cr
= &
{\kappa_5^2 L_{\rm eff} \over 32 \pi T_{\rm eff}}
        {\omega \over
         \left(1- \e^{-{\omega\over 2 T_L}} \right)
         \left(1- \e^{-{\omega\over 2 T_R}} \right)
        }
        \left( \omega^2 + 16 \pi^2 T_L^2 \right) \ .
}}
The values of the parameters in the effective string model
have been fixed in \refs{\juan,\MSI,\CGKT},
\eqn\parameters{
\kappa_5^2 L_{\rm eff}= 4 \pi^3 r_1^2 r_5^2\ , \qquad
T_{\rm eff}= {1\over 2 \pi r_5^2}\ ,
}
where
$$ r_1^2\equiv  \hat Q\ , \qquad
r_5^2\equiv  \hat P\ .
$$
Note that this effective string tension is the tension of the
D-string divided by $n_5$, the number of 5-branes. This value of
the tension is necessary for agreement with the entropy of near-extremal
5-branes \juan, as well as for the agreement of the fixed-scalar
cross-section for
$r_1=r_5$ \CGKT.
In this paper we will show that it also leads to agreement
of the absorption cross-sections for the  scalars $\eta_i$ and $\xi_i$.

Using \llr, \parameters, and the detailed balance,
we find that the absorption cross-section
for $\eta_i$ is
\eqn\offdiag{
\sigma_{\eta}(\omega) = { \pi^3  r_1^2 r_5^4 \over 4}
{\omega \left (\e^{\omega\over T_H} - 1 \right ) \over
\left (\e^{\omega\over 2 T_L} - 1\right )
\left (\e^{\omega\over 2 T_R} - 1\right ) }
(\omega^2 + 16 \pi^2 T_L^2 )
\ .
}
After analogous steps, the absorption cross-section for $\xi_i$
is found to be
\eqn\otheroffdiag{
\sigma_\xi(\omega) = { \pi^3  r_1^2 r_5^4 \over 4}
{\omega \left (\e^{\omega\over T_H} - 1 \right ) \over
\left (\e^{\omega\over 2 T_L} - 1\right )
\left (\e^{\omega\over 2 T_R} - 1\right ) }
(\omega^2 + 16 \pi^2 T_R^2 )
\ .
}

In the next section we will check these greybody factors against
semi-classical effective field theory  calculations.
We will need the following expressions for
the temperatures \MSI,
\eqn\temps{
T_L = {r_0 e^ \sigma \over 2 \pi r_1 r_5}\ ,
\qquad T_R = {r_0 e^ {- \sigma} \over 2 \pi r_1 r_5}\ ,
\qquad {2\over T_H} = {1\over T_L} + {1\over T_R}\ ,}
where
 $\sigma$ is defined by
$$ r_n^2= r_0^2 \sinh^2 \sigma
\ , \ \           \qquad r_n^2\equiv  \hat Q_K\ .       $$
This may be solved with the result,
$$ e^{\pm 2\sigma} = 1+ {2\over r_0^2} (r_n^2 \pm Q)\ .
$$
Under $Q_K \rightarrow -Q_K$, we therefore find that
$\sigma \rightarrow -\sigma$,
which implies that $T_L$ and $T_R$ are interchanged.
This transformation reverses the momentum flow along the
string, so that the operators of dimension
$(1,2)$ and $(2,1)$, and therefore $\xi_i$ and
$\eta_i$, are interchanged.
The classical equations for $\xi_i$ and $\eta_i$, \kgoo, \masses,
are also interchanged under
$Q_K \rightarrow -Q_K$. This is the first, and very important,
consistency check between the effective string and the semi-classical
descriptions of the intermediate scalars.

\newsec{Comparison with Semiclassical Greybody Factors}

In this section we   carry out a number of calculations which
indicate agreement, at least in various limits,
between the semi-classical cross-sections and
those in the effective string model. First we
discuss the case $Q_K=0$ where the classical calculation is the easiest.
Then we address various limits of the $Q_K>0$ case.

\subsec{ $Q_K=0$}

Here we consider the case $r_n^2=0$ (i.e. $Q_K=0$), where
$\eta_i$ and $\xi_i$ satisfy identical equations \kgoo,\kgooo.
Since here $T_L=T_R$, the two effective string greybody factors are also
the same, and they will turn out to be identical to the
semi-classical ones.

The non-extremal equation satisfied by both $\eta_i$ and $\xi_i$
is (for $r_0 \ll r_1, r_5$)
\eqn\ExactNonExtreme{
\bigg[    hr^{-3} {\del_r }\big
( h r^3 {\del_r} \big ) +  f(r)\omega^2 - 3 h{r_1^4 \over r^2
(r_1^2 + r^2)^2}
+ h {r_0^2 r_1^2 (r_1^2 + 2r^2)\over r^4 (r_1^2 + r^2)^2}
\bigg] R = 0 \ ,
}
where  we set $\eta_i,\xi_i =R(r) e^{i\o t}$,  and
$$ h(r)= 1 - {r_0^2\over r^2}\ , \qquad
f(r)= \big(1 + {r_1^2\over r^2}\big)
\big(1 + {r_5^2\over r^2}\big)
\ .$$

In the near region ($r\ll r_1, r_5$) we find,
in terms of the variable $z=h(r)$,
\eqn\one{
\bigg[ z{\del_z }( z{\del_z }) +
 D+{C\over (1-z)}+{E\over (1-z)^2}\bigg ]R=0 \ ,
}
where
$$D=-{1\over 4}\ ,\qquad
C= {\omega^2 r_1^2 r_5^2 \over 4 r_0^2} + 1\ , \qquad
E=-{3\ov 4}\ .$$
This may be reduced to a hypergeometric equation by
a substitution of the form
\eqn\two{
R=z^\alpha(1-z)^\beta F(z) \ .
}
After some algebra we find that, if $\alpha$ and $\beta$
satisfy
$$E+\beta(\beta-1)=0\ , \qquad \alpha^2+D+C+E=0\ ,
$$
then the equation for $F(z)$ becomes
\eqn\hyperg{
z(1-z){d^2 F\over dz^2}+
[(2\alpha+1) (1-z)-2\beta z] {d F\over dz}
-[(\alpha+\beta)^2+D] F=0
\ ,}
which is the hypergeometric equation.
In general, the solution to
\eqn\genhyper{
z(1-z){d^2 F\over dz^2}+
[C- (1+A+B) z] {d F\over dz}
-AB F=0\ ,
}
which satisfies $F(0)=1$, is the hypergeometric function $F(A,B;C;z)$.
Thus, the solution in the inner region is
\eqn\inner{
R_I= z^\alpha (1-z)^\beta F(\alpha+\beta+ i\sqrt D,
\alpha+\beta- i\sqrt D; 1+ 2\alpha; z)
\ ,
}
where
$$\beta=-{1\ov 2}  \ , \qquad
\alpha=-i {\omega r_1 r_5\over 2 r_0}=-i {\omega\over 4\pi T}\ .
$$
In the last equation we used the fact that, for $r_n=0$,
$$ T=T_L= T_R= T_H= {r_0\over 2\pi r_1 r_5}
\ .$$
Using the asymptotics of the hypergeometric functions for
$z\rightarrow 1$, we find that,
for large $r$,
$$ R_I\  \rightarrow
\ {r\over r_0} E\ ,
$$
where
$$ E=
 {\Gamma \left (1- i {\omega\over 2\pi T}\right )\over
\Gamma \left (2- i {\omega\over 4\pi T} \right )
\Gamma \left (1- i {\omega\over 4\pi T} \right ) }
\ .$$

In the middle region ($r_0\ll r\ll 1/\omega$) the approximate solution is
$$ R_{II} \approx  E \ {r_1\over r_0} \big (1+ {r_1^2\over r^2}\big)^{-1/2}\ .
$$
In the outer region,
the dominant solution, which matches to the asymptotic form in region
II, is
$$R_{III}=2 A\rho^{-1} J_1 (\rho)\ , \qquad \rho= \omega r
\ .$$
By matching we find that
$$ A= E \ {r_1\over r_0}\ . $$

The absorption cross-section may
now be obtained using the method of fluxes (see, e.g.,
 \refs{\MSI,\CGKT} and references therein).
The flux per unit solid angle is
\eqn\Flux{
   {\cal F} =  {1\over 2i} (R^* h r^3 \partial_r R - {\rm c.c.}) \ .
}
 The absorption probability is the ratio of the incoming flux at
the horizon to the incoming flux at infinity,
$$
   P= {{\cal F}_{\rm horizon} \over  {\cal F}_\infty^{\rm incoming}}
    = {\pi \omega^3\over 2} r_1 r_5 r_0
\ |A|^{-2} \ .
$$
The absorption cross-section is related to the s-wave absorption
probability by
$$
   \sigma_{\rm abs} = {4 \pi\over \omega^3} P=
{2\pi^2 r_0^3 r_5\over r_1} |E|^{-2}
\ .
$$
Thus,
$$ \sigma_{\rm abs}= {2\pi^2 r_0^3 r_5\over r_1} x (1+x^2)
{e^{2\pi x}+1\over e^{2\pi x}-1} \ ,
$$
where
$$ x= {\omega\over 4\pi T}= {\omega r_1 r_5\over 2 r_0}\ .
$$
It follows that
\eqn\class{
\sigma_{\rm abs}= {\pi^3 \over 4} r_1^2 r_5^4
{e^{\omega\over 2T}+1\over e^{\omega\over 2T}-1}
\omega (\omega^2 + 16 \pi^2 T^2)
\ .
}
This is in {exact}  agreement with the cross-sections
\offdiag\ and \otheroffdiag\ derived in the
effective string model! In particular, the agreement of the
overall normalization provides new evidence in favor of the effective
string tension \parameters\ being given by
the D-string tension divided by $n_5$.

\subsec{The $\xi_i$ cross-section for $Q_K>0$}

The scalar $\xi_i$ has the fluctuation equation \kgoo\
with the effective mass term $M_\xi= M_{\l- \n}+M_+$.
We will try to solve for the cross-section exactly in the regime
where $r_0\ll r_n\ll r_1, r_5$, so that $T_R\ll T_L$. We will
take $\o/T_R$ to be of order 1. Hence we should be
able to find the dependence of the cross-section on $\o/T_R$,
which is a test of the greybody factor dependence.

We will match the approximate solutions in several regions.
First, consider the inner region, $ r\ll r_n$.
Here the effective mass is approximately
$ {8h\over r^2} \ , $
and the equation becomes
$$
\bigg[ hr^{-3}\del_r(hr^3\del_r ) +
{\o^2 r_1^2 r_5^2 r_n^2\over r^6}-{8(r^2-r_0^2)\over
r^4} \bigg] R =0
\ .$$
In terms of the variable $z=1-{r_0^2\over r^2}$,
$$
\bigg[ z\del_z(z\del_z) +
{\o^2 r_1^2 r_5^2 r_n^2\over 4r_0^4}-{2z\over (1-z)^2}\bigg] R
=0
\ .$$
This equation has the same form as \one\ with
$$
D={\o^2r_1^4r_n^2\over 4r_0^4}\ ,\qquad C=2\ ,\qquad E=-2\ .
$$
We will again use the  substitution \two, where now
$$
\eqalign{&
E+\b(\b-1)=0\ \rightarrow\ \  (\b-2)(\b+1)=0\ ,\cr
& \a^2+D+C+E=0\ \rightarrow\ \  \a=- i{\o r_1 r_5 r_n\over 2r_0^2}
\ . }
$$
For $r_0\ll r_n$, we have
\eqn\tempright{
T_R\approx {r_0^2\over 4\pi r_1 r_5 r_n}
\ .}
Thus,
\eqn\formal{
\a=- i {\o\over 8\pi T_R}
\ .}
We also choose $\beta=-1$. Hence,
the solution is
$$
R_I=z^\a(1-z)^{-1} F(-1+\a+i\sqrt{D}, -1+\a-i\sqrt{D};1+2\a;z)
\ .$$
Away from the horizon, i.e. as $z\rightarrow 1$,
$$
R_I\rightarrow {r^2\over r^2_0}{2\over 1-i{\o\over 4\pi T_R}}
\equiv K_1  {r^2\over r^2_0} \ .
$$

Now we discuss the region
$ r_0 \ll r \ll r_1, r_5$.
Here we may drop the $\o^2$ term, and also set $h=1$.
The equation becomes
$$\left[r^{-3}\del_r(r^3\del_r)-{3r^4+8r_n^2r^2+8r_n^4 \over
r^2(r^2+r_n^2)^2}\right]R=0
\ .$$
Substituting $t=1+{r_n^2\over r^2}$, we find
$$
4\del_t^2 R - {3-8t+8t^2\over (1-t)^2t^2}R=0
\ .$$
One may check that the solutions are
$$
[t(t-1)]^{-1/2} \qquad{\rm and}\qquad t^{3/2}(t-1)^{-1/2}(3-2t)
\ .
$$
To match to the near-horizon solution we choose
$$
R_{II}=K_2[t(t-1)]^{-1/2}\ ,\qquad K_2={r_n^2\over r_0^2}K_1
\ .
$$
For large $r$ the solution approaches
$$
R_{II}=K_2{r\over r_n}
\ .$$

In the intermediate region, $r_n\ll r\ll 1/\o$, we have
\eqn\intereq{
\left [r^{-3}\del_r(r^3\del_r)-{3 r_1^4\over r^2(r^2+r_1^2)^2}
\right ]R=0\ ,
}
with the solution
$$
R_{III}=K_2{r_1\over r_n}\left(1+{r_1^2\over r^2}\right)^{-1/2}
\ .$$

In the far region, $r\gg r_1$, we have
$$
[r^{-3}\del_r(r^3\del_r)+\o^2]R=0
$$
with the solution
\eqn\far{
R_{IV}=2A(\o r)^{-1}J_1(\o r)
\ .}
Matching the solutions, we find
$$
A=K_2{r_1\over r_n}=K_1{r_1r_n\over r_0^2}
\ .$$

The absorption cross-section is given by
\eqn\absc{
\s_{\xi}={4\pi\over \o^3}{\pi\o^3\over 2}
r_1 r_5 r_n|A|^{-2}
={\pi^2 r_0^4 r_5\over 2 r_1 r_n}
\left(1+{\o^2\over 16\pi^2T_R^2}\right)
\ .}
Note that this is exact in $\o/T_R$.
To compare this with the effective string result, we
take the limit $\o/T_L \rightarrow 0$ in \otheroffdiag.
Using \tempright\ and
$$ T_L \approx {r_n \over \pi r_1 r_5}\ ,
$$
we find exact agreement of the two greybody factors.

\subsec{The $\eta_i$ cross-section for $Q_K>0$}

Let us now consider  the scalar
$\eta_i$, which satisfies \kgoo\ with the effective mass
$M_\eta= M_{\lambda- \n}+ M_-$.
We will again solve for the cross-section exactly in
the limit $r_0\ll r_n\ll r_1$.

In the inner region, $r\ll r_n$,
we may ignore the mass term.
The equation is
$$
\bigg[ hr^{-3}\del_r(hr^3\del_r) + {\o^2r_1^2 r_5^2 r_n^2\over r^6}\bigg] R =0
$$
which may be written as
$$
\bigg[ z\del_z(z\del_z) +  {\o^2
r_1^2 r_5^2 r_n^2\over 4r_0^4}\bigg]  R =0
\ .$$
The solution is
$$
R_I=z^\a\
$$
with $\a$ given in \formal.
Away from the horizon, i.e. for $z\rightarrow 1$,
$ R_I\rightarrow 1 $.

In the region $ r_0 \ll r \ll r_1, r_5$
the approximate equation is
$$
\left[r^{-3}\del_r(r^3\del_r)-{3r^2 \over (r^2+r_n^2)^2}\right]R=0
\ ,
$$
which may be recast as
$$
4\del_t^2R - {3\over (t-1)^2t^2}R=0
\ .$$
One may check that the solutions are
$$
t^{3/2}(t-1)^{-1/2} \qquad{\rm and}\qquad t^{-1/2}(t-1)^{-1/2}(2t-1)
\ .
$$
To match to the near horizon solution we pick
$$
R_{II}={1\over 2}t^{-1/2}(t-1)^{-1/2}(2t-1)
\ .$$
For large $r$, $ R_{II}\rightarrow {r\over 2r_n}$.

In the intermediate region, $r_n\ll r\ll 1/\o$,
the equation is again given by \intereq. The
solution matching $R_{II}$ is
$$
R_{III}={r_1\over 2r_n}\left(1+{r_1^2\over r^2}\right)^{-1/2}
\ .$$

In the far region, we again find a solution of the form \far.
Matching the solutions, we find $ A={r_1\over 2r_n}$.
The absorption cross-section is given by
\eqn\newcross{
\s_{\eta}=2\pi^2 r_1 r_5 r_n|A|^{-2}
=8\pi^2 {r_n^3 r_5\over r_1}\ .
}

This should be compared with the $\omega/T_L\rightarrow 0$
limit of the effective string greybody factor, \offdiag.
Once again, we find exact agreement!

\subsec{The Low Temperature limit}

In this section we analyze the
$\o \gg T_L,T_R$ limit of the greybody factors.
In the inner region we ignore the mass term
since it is smaller than the $\o^2$ term,
$$
\bigg[  hr^{-3}\del_r(hr^3\del_r)
 +{\o^2r_1^2 r_5^2\over r^4}\big( 1+{r_n^2\over
r^2}\big)\bigg]  R =0
\ .$$
The solution is
$$
R_I=z^{-i{(a+b)/ 2}} F(-ia,-ib,1-ia-ib,z)\ ,
$$
with
$$
a= {\o\over 4\pi T_L} \ ,
\qquad b= {\o\over 4\pi T_R}   \ .
$$
As $z\rightarrow 1$,
$$
R_{I}\rightarrow E={\G(1-ia-ib)\over \G(1-ib)\G(1-ia)}
\ .$$

In the region $r_1, r_5\gg r\gg r_n$,
the equation is
$$
\left[r^{-3}\del_r(r^3\del_r)+
{\o^2 r_1^2 r_5^2 \over r^4}-{3 \over r^2}\right]R=0
\ .$$
Substituting $u={\o r_1 r_5\over r}$, we get
$$
\bigg[ u\del_u ({u\inv}\del_u)  - 1+ {3\over u^2}\bigg] R=0
$$
with the solution
$$
R_{II}={K_3\over r}\ N_2 \left( {\o r_1 r_5\over r}\right)
\ .
$$
To match to the near-horizon solution we first
introduce an auxiliary function $\R$  satisfying
$$
\left[r^{-3}\del_r(r^3\del_r)+{\o^2 r_1^2 r_5^2\over r^4}\right]\R=0
\ , $$
so that
$$
\R ={\psi\over r}\ N_2\left({\o r_1 r_5\over r}\right)
\ .$$
Matching $\R$ and $R_{II}$ for small $r$, we find $\psi=K_3$.
Matching  $\R$ and $R_{I}$ for large $r$, we find
$$
\psi={\pi\o r_1 r_5\over 2}E=K_3
\ .$$
For large $r$,
$$
R_{II}\rightarrow K_3{4r\over \pi\o^2 r_1^2 r_5^2}
\ .$$

In the intermediate region
the equation is again given by \intereq, and now
the solution is
$$
R_{III}=K_3{4\over \pi\o^2 r_1 r_5^2}
\big(1+{r_1^2\over r^2}\big)^{-1/2}
\ .$$
In the far region,
$r\gg r_1, r_5$, the solution is again of the form
\far.
Matching the solutions, we find
$$
A=K_3{4\over \pi\o^2 r_1 r_5^2}={2\over \omega
r_5}{\G(1-ia-ib)\over \G(1-ib)\G(1-ia)}
\ .$$
The absorption cross-section is given by
$$
\s_{\rm abs}=2\pi^2 r_1 r_5 \sqrt {r^2_n+ r^2_0}
\ |A|^{-2}
={\pi^2\over 4}r_1 r_5^3 r_0 \cosh \sigma {\o^3\over T_L+T_R}
\ , $$
where we have used the condition $\o\gg T_L,T_R$
to simplify the exponentials.
Since the temperatures  \temps\ satisfy
$$ T_L+T_R = {r_0 \cosh \sigma \over \pi r_1 r_5}\ ,
$$
we finally have
\eqn\newnew{
\s_{\rm abs}= {\pi^3 \over 4} r_1^2 r_5^4 \omega^3
\ . }
Note that the effective string greybody factors for both scalars,
\offdiag\ and \otheroffdiag,
exactly agree with this for $\o\gg T_L,T_R$. Thus, this is another point
of agreement between the semi-classical gravity and the effective string.

\newsec{Conclusions}

A remarkable feature of the charged supersymmetric black holes is the
variety of physically different behaviors exhibited by
scalar fields. The minimally coupled and the fixed scalars have been
thoroughly analyzed in earlier work, and this paper is devoted to a yet
different type of scalars, which we call intermediate.

In the effective string models the physical differences between scalars
are due to the different operators they couple to.
Indeed, the leading coupling of the minimal scalars is to operators
of dimension $(1,1)$, while that of the fixed scalars is to operators
of dimension $(2,2)$. In \CGKT\ it was observed that the intermediate
scalars couple to chiral operators of dimension $(1, 2)$ and $(2, 1)$.
The main achievement of the present work is to  discover
 a surprising
mixing between the  intermediate scalars from the NS-NS and the R-R
sectors. Thus, we find two different intermediate scalars; one of
them appears to couple to a dimension $(1, 2)$ operator, and the
other to a dimension $(2, 1)$ operator.
In the absence of a
momentum flow along the string (the Kaluza-Klein charge),
the string theory is non-chiral, and
there is no physical difference between these two operators.
In this regime we indeed find that both intermediate scalars
satisfy the same classical equation and, therefore, have identical
semi-classical
greybody factors, which turn out to agree with the effective string
model exactly.

When the momentum flow is present ($Q_K\neq 0$), 
the effective string model is chiral,
and the two intermediate scalars have different greybody factors,
\offdiag\ and \otheroffdiag. Remarkably, now the two classical
fluctuation equations \kgoo\ are different: their near-horizon behavior
jumps when $Q_K$ is turned on. This jump works in precisely the right
way for the semiclassical greybody factors to agree, at least
in certain regimes, with the effective string ones.
In general we find that, both in the effective string and in the 
semi-classical approaches, the two intermediate scalars are 
interchanged by reversing the momentum flow, $Q_K\rightarrow
- Q_K$.

We have tested
the greybody factors predicted by the effective string model
against the  semi-classical calculations in the following
regimes,
$$
\eqalign{& A. \qquad\qquad {\omega\over T_L}= {\omega\over T_R}
\sim O(1)\ ;\cr
& B. \qquad\qquad {\omega\over T_L} \ll 1\ ,\qquad
{\omega\over T_R} \sim O(1)\ ;\cr
& C. \qquad\qquad {\omega\over T_L} \gg 1\ ,\qquad
{\omega\over T_R} \gg 1\ ,
}
$$
finding complete agreement.
Unfortunately, we have not been able to extract the semi-classical
absorption cross-sections as functions of ${\omega\over T_L}$ and
${\omega\over T_R}$ in general.
For this reason, further analysis of the classical equations \kgoo,
and comparison with the effective string greybody factors, \offdiag\
and \otheroffdiag, is desirable. This could provide a further sensitive
test of the effective string model.

\newsec{Acknowledgements}

We are grateful to  J. Maldacena for useful discussions.
 The  work of I.R.K.
was supported in part by DOE grant DE-FG02-91ER40671,
the NSF Presidential Young Investigator Award PHY-9157482, and the
James S.{} McDonnell Foundation grant No.{} 91-48.
The work of A.R. was supported in part by the Department of Energy
under contract no. DE-AC03-76SF00515.
A.A.T. is grateful to the Institute of Theoretical Physics of
SUNY at Stony Brook
for hospitality during the final stage of this work
and  acknowledges also the support
 of PPARC and  the European
Commission TMR programme grant ERBFMRX-CT96-0045.

\appendix{A}{Equations for
intermediate scalars: the $D=6$ perspective}
The $D=5$ black hole background \met,\torr\
may be viewed as a dimensional reduction of a boosted
 solitonic black string  solution  in $D=6$.
 The  equations   for small perturbations near this
 background can thus  be derived by expanding the type IIB action
 reduced to  6 dimensions, assuming that the $D=6$ fluctuations do not depend
 on the string direction $x^5$.
 This method of derivation  clarifies
  the reason behind the mixing of the fields $A^i_5$
 and $B_{5i}$. In $D=6$ they appear as the  $M=5$
 components of the $D=6$
 vectors: the KK one,  $A^i_M$, and the RR one, $B_{Mi}$ ($i=1,2,3,4; \
 M=0,1,2,3,4,5$).
 Another  conceptual  advantage of the $D=6$ approach is
that it enables us to include the dependence on the Kaluza-Klein
   charge, $Q_K$, simply by a coordinate transformation
    (a finite Lorentz boost in the string direction)
of the non-extremal case with $Q_K=0$.\foot{The crucial point is that
   the dependence of the  non-extremal $D=6$  solution  on $Q_K$  can be
induced by a finite boost,
  which is not a symmetry of  the black string  background for
  $r_0 \not=0$.}

Let us  first consider   the  $Q_K=0$ case.
The $D=6$ black string has a trivial  dilaton background,
$\p_6=0$, so that there is no
difference between the Einstein and the string metric.
 The KK scalar matrix   is
$$G_{ij} = e^{2\n} \d_{ij}\ ,\qquad
\ \sqrt G = e^{4\n}\ ,$$
and the metric is
$$
ds^2_6 = (H_1H_5)^{-1/2} (-h dt^2 + dy_5^2) + (H_1H_5)^{1/2} (h\inv dr^2
+ r^2 d\Omega_4^2) \ ,
$$
$$ e^{2\n} = (H_1/H_5)^{-1/2} \ ,  \
\ \ \  H_{0r5} = 2Qr^{-3}  H_1^{-2}, \ \ \ \
  \sqrt g  e^{4\n} H^{0r5} = 2 Q\  .$$
The R-R antisymmetric tensor field strength, $H_{MNK}$, also has
 the magnetic (5-brane) components  which will not
couple to the fluctuation fields  we are interested in.

The relevant part of the $D=6$ action that
governs  small  fluctuations of the vector
fields $A^i_M$ and $B_{Mi}$,
 which have trivial background values,  is
$$
S_6 =
 {1 \ov 2 \k_6^2} \int d^6 x \sqrt g \big[   -
 \fourth G_{ij}  F^{i}_{MN}  F^{j MN}
- \fourth   \sqrt G G^{ij}
 H_{MN i } H^{MN}_{\ \ \ \ j}
-\  {1\ov 12} \sqrt G H^{MNK}H_{MNK} \big]  $$
$$ ={1 \ov 2 \k_6^2} \int d^6 x \sqrt g \big[   -
 \fourth  e^{2\n}   F_{iMN}  F^{i MN}
- \fourth   e^{2\n}
 H_{MN i } H^{MNi}$$
  \eqn\onn{   \   + \  \fourth  e^{4\n}
 H^{MNK} ( A^i_M H_{NK i} +  B_{M i} F^i_{NK})
 + ...  \big] \ ,   }
where
$$H_{MNK} = \del_M B_{NK} - \ha A^i_M H_{NK i}-
\ha B_{M i} F^i_{NK}
+ {\rm cyclic} \ , \ \ \   H_{MNi} = \del_M B_{Ni} + {\rm cyclic} \ . $$
Here $B_{MN}$ and $B_{Mi}$ are  equal to  the corresponding
components
of the $D=10$ R-R 2-tensor, up to terms proportional to the
KK vector $A^i_M$ whose precise form  will not be important.

The term linear in $H_{MNK}$ mixes the two vector perturbations in the
string background
(the two terms in the  bracket
give equivalent contributions because
$H_{MNK} $ has an on-shell value).
It is only  the electric part of the $H_{MNK}$ background
that  contributes to the equations  for the
$A^i_5,\  B_{i5}$  components  we are interested in.

 It is crucial that the  background factors  in the
 kinetic terms for
 $A^i_5$ and $B_{i5}$  are the same.
This is a consequence of the R-R nature of $B_{5i}$ and is not true
for its NS-NS counterpart.
 As a result,  one may diagonalize the action introducing
 $$A^{i + }_M =
{1\ov \sqrt 2 } ( A^i_M  +  B_{M i}  ) \ , \ \ \ \
A^{i - }_M= {1\ov \sqrt 2 }  ( A^i_M -  B_{M i}  ) \ ,
\ \ \ \ \
$$
so that the scalars in \qqq\ are  $A^{i - }_5=\xi_i  , \
 \  A^{i + }_5=\eta_i $. The action becomes
$$
S_6= S_+ + S_-  =
  \int d^6 x \sqrt g \bigg[   -   \fourth e^{2\n} F^{+i}_{MN}  F^{+i  MN}
  + \fourth e^{4\n} H^{MNK} A^{+i}_M  F^{+i}_{NK}   $$
\eqn\tite{
  -  \ \fourth e^{2\n} F^{-i}_{MN}  F^{-i  MN}
  -   \fourth e^{4\n} H^{MNK} A^{-i}_M  F^{-i}_{NK}  \bigg]   \ .   }
Keeping only the relevant electric components  we get,
$$
 S_+ =
  \int d^6 x \sqrt g\ e^{2\n} \bigg[  \hal   F^{+i}_{0r}  F^{+i  0r }
  +  \hal  F^{+i}_{05}  F^{+ i 05 }  - \hal F^{+}_{5r}  F^{+i  5r }
  $$ \eqn\onn{
  +\  \hal e^{2\n} H^{05r} (A^{+i}_5  F^{+i}_{r0}  + A^{+i}_0 F^{+i}_{5r}
   + A^{+i}_r  F^{+i}_{05})
   \bigg]   \ , }
and similarly for $S_-$.
 Since we  assume that all the fields
  do not depend on $x^5$ (but, in fact, depend only on $r$ and $t$),
  and that the $H_{MNK}$
  background is on-shell, the last term is
  equal to $e^{2\n}  H^{05r} A^{+i}_5  F^{+i}_{r0}$,
up to a total derivative.

  $S_+$  may be viewed as an action  for  a 2d vector $A^+_a$ $(a=0,r)$
  coupled to a scalar $A^{+i}_5= \eta_i$.
 To establish a correspondence with the
  $D=5$ picture, it is natural to  integrate out
  the $(0,r)$ components of the vector $A^+_M$,
  which is equivalent in the present context
  to solving for $F^{+i}_{r0}$.
  As a result,
  \eqn\onn{
 S_+ =
  \int d^6 x \sqrt g e^{2\n}  \bigg[
      - \hal \del_a \eta_i\del^a \eta_i
  -  \hal e^{4\n} H^{05r} H_{05r} \eta_i \eta_i   \bigg] \ ,  }
  so that we get the scalar equation
  for $ \eta_i  $ with the mass term determined by
  the  electric part of the $H_{MNK}$ background
  and originating from the KK vector --
   R-R vector   mixing.
   This  is  the same
    mass term for the (rescaled) $\eta_i$ as
    found in the  $D=5$ approach of Section 2 for the case of $Q_K=0$
    (see \ait,\kgoo).

To  find the perturbation  equations that
include  the dependence on  the third charge,
$Q_K$,  we may use  the fact that the action is invariant under
reparametrizations. One may  either  apply a boost
in the $x^5$ direction to  the background  fields
or  keep the background    unchanged,
but instead transform the vector components $A^{\pm i}_M$, $M=0,5$.
The  boost  interpretation of the $Q_K$ dependence
is  manifest in $D=6$
 before  one  integrates  out
the $0r$ component of the  vector field strength
 (the boost `mixes' $F_{0r}, F_{05}, F_{r5}$).
The presence of the $Q_K$-dependent   cross-term in \gins\
   is  understood  from the  $D=6$ perspective to be a  consequence
   of  the mixing  between $A^i_M$ and $B_{Mi}$
occurring already for $Q_K=0$, and of the fact that
 a non-zero boost
creats an  extra term  in the action  which is linear in $F_{05}$.
As a result, one finds the  same equations \kgoo\
as  obtained  in  the $D=5$  approach.

\listrefs
\bye